\documentclass[prd,preprintnumbers,nofootinbib,tightenlines,superscriptaddress]{revtex4}
\pdfoutput=1
\usepackage{graphicx,epstopdf}
\usepackage{amssymb}
\usepackage{amsmath,mathrsfs,verbatim}
\usepackage[usenames, dvipsnames]{color}
\usepackage{slashed}
\usepackage{times}
\usepackage{latexsym}

\usepackage{floatrow}

\usepackage{longtable}
\usepackage[utf8]{inputenc}
\usepackage{hyperref}

\begin{document}

\title{Astrophysical probes of inelastic dark matter with a light mediator}

\author{Gerardo Alvarez}
\email[]{gerardo.alvarez@email.ucr.edu}
\affiliation{Department of Physics and Astronomy, University of California, Riverside, California 92521, USA}
\author{Hai-Bo Yu}
\email[]{hai-bo.yu@ucr.edu}
\affiliation{Department of Physics and Astronomy, University of California, Riverside, California 92521, USA}

\date{\today}

\begin{abstract}

We consider an inelastic dark matter model, where a fermion is charged under a broken U(1) gauge symmetry, and introduce a tiny Majorana mass term to split the fermion into two states with the light one being a dark matter candidate. If the gauge boson is light, it can mediate both elastic and inelastic dark matter self-interactions in dark halos, leading to observational consequences. Using a numerical technique based on partial wave analysis, we accurately calculate the elastic and inelastic self-scattering cross sections. We assume a thermal freeze-out scenario and fix the gauge coupling constant using the relic density constraint. Then, we focus on six benchmark masses of dark matter, covering a wide range from $10~{\rm MeV}$ to $160~{\rm GeV}$ and map parameter regions where the elastic scattering cross section per unit mass is within $1~{\rm cm^2/g}\textup{--}5~{\rm cm^2/g}$, favored to solve small-scale issues of cold dark matter. If the heavy state can decay to the light state and a massless species, the inelastic up-scattering process can cool the halo and lead to core collapse. Taking galaxies with evidence of dark matter density cores, we further derive constraints on the parameter space. For dark matter masses below $10~{\rm GeV}$, the mass splitting must be large enough to forbid up scattering in the dwarf halo for evading the core-collapse constraint; while for higher masses, the up-scattering process can still be allowed. Our results show astrophysical observations can provide powerful tests for dark matter models with large elastic and inelastic self-interactions. 

\end{abstract}


\maketitle

\section{Introduction}
\label{sec:intro}

It is well established there is a non-luminous component in the universe, called dark matter. Since its influence has, so far, only been observed to be gravitational in nature, the particle properties of dark matter are still largely elusive. Self-interacting dark matter (SIDM) is a class of particle physics models in which dark matter particles are assumed to have a large self-scattering cross section~\cite{Spergel:1999mh,Kaplinghat:2015aga}, see~\cite{Tulin:2017ara} for a recent review. N-body simulations of structure formation demonstrate strong dark matter self-interactions can lead to heat transfer in the halo and allow its inner region to thermalize~\cite{Dave:2000ar,Vogelsberger:2012ku,Zavala:2012us,Peter:2012jh,Rocha:2012jg,Robertson:2016qef,Kummer:2019yrb}. Recently, it has been shown that SIDM can explain a number of long-standing puzzles in astrophysics presented in the prevailing cold dark matter theory, such as diverse galaxy rotation curves of spiral galaxies in the field~\cite{Kamada:2016euw,Creasey:2016jaq,Ren:2018jpt,Kaplinghat:2019dhn}, dark matter distributions in satellite galaxies in the Milky Way~\cite{Vogelsberger:2012ku,Valli:2017ktb,Nishikawa:2019lsc,Kaplinghat:2019svz,Sameie:2019zfo,Kahlhoefer:2019oyt}, and shallow inner dark matter density profiles in galaxy clusters~\cite{Newman:2012nw,Kaplinghat:2015aga}. SIDM also inherits all the success of cold dark matter in explaining large-scale structure of the universe~\cite{Ade:2015xua} and many important aspects of galaxies~\cite{Springel:2006vs,TrujilloGomez:2010yh}.

Many SIDM models assume there is only one dark matter state and a light force carrier mediates elastic dark matter self-scattering in the halo, see, e.g.,~\cite{Tulin:2017ara}. In this case, the initial and final states in the collisions are the same, and they only redistribute energy of dark matter particles as the halo as a whole does not lose energy. More recently, there is growing interest in considering particle physics realizations of SIDM with multiple states. For example, to avoid strong bounds from direct detection experiments~\cite{Kaplinghat:2013yxa,Ren:2018gyx, Akerib:2018hck,Aprile:2019xxb}, Refs.~\cite{Zhang:2016dck,Blennow:2016gde} propose an inelastic SIDM model, where there are two dark states and they differ by a small mass splitting~\cite{Han:1997wn,TuckerSmith:2001hy}. One can adjust the splitting to kinematically forbid transitional up-scattering in nuclear recoils, but still allow strong elastic self-interactions between two light states in the halo. In addition, Ref.~\cite{Schutz:2014nka} studies dark matter self-interactions in the exciting dark matter model~\cite{Finkbeiner:2007kk,ArkaniHamed:2008qn,Loeb:2010gj,Finkbeiner:2014sja}, where dark matter collisions can produce a heavy state that subsequently decays back to the light one and a standard model particle. More generally, if the SIDM candidate is made of composite states, such as dark atoms~\cite{Mohapatra:2000qx,Mohapatra:2001sx,Kaplan:2009de,Khlopov:2010ik,Cline:2012is,CyrRacine:2012fz,Cline:2013pca,Foot:2014mia,Foot:2014uba,Boddy:2016bbu,Buckley:2017ttd} and strongly-coupled particles~\cite{Frandsen:2010yj,Frandsen:2011kt,Boddy:2014qxa,Braaten:2018xuw}, it is natural to expect inelastic excitations during dark matter collisions in dark halos. 

In this paper, we consider an inelastic SIDM model and study its astrophysical implications. It assumes that a Majorana mass term induces a small mass splitting between two fermionic dark matter states and they interact with a U(1) gauge boson. We assume the gauge boson is light and develop a numerical method to calculate both elastic and inelastic dark matter self-scattering cross sections. After imposing the relic abundance constraint on the gauge coupling constant, we focus on benchmark dark matter masses, which cover a wide range from $10~{\rm MeV}$ to $160~{\rm GeV}$, and search for parameter regions where the elastic self-scattering cross section per unit mass ($\sigma_V/m_\chi$) satisfies $1~{\rm cm^2/g}\leq \sigma_V/m_\chi\leq5~{\rm cm^2/g}$, as favored by observations on galactic scales~\cite{Tulin:2017ara}. Our work is a natural and simple extension to the minimal SIDM model that contains only one dark matter state, but calculations of the self-scattering cross sections in the current model are much more challenging than the minimal one~\cite{Feng:2009hw,Tulin:2013teo}. In addition, we explore a broader mass range for both dark matter and mediator particles, compared to earlier studies~\cite{Blennow:2016gde}. As we will show, for the dark matter mass below $\sim1~{\rm GeV}$, inelastic up scattering dominates over elastic one if the former is kinematically open in the dark halo. This has important implications for constraining the parameter space. 

We further consider the endothermic up-scattering process of dark matter particles and its influence on halo evolution and  inner halo structure. If the heavy state decays back to the light state by releasing a massless species, the SIDM halo profile can become cuspy again, because the dissipative self-interactions can cool the inner halo and speed up the onset of core collapse~\cite{Essig:2018pzq,Choquette:2018lvq}. Using dwarf galaxies that show density cores, Ref.~\cite{Essig:2018pzq} derives constraints on parameters that characterize the cooling rate of dissipative dark matter collisions. In this work, we will take the results in~\cite{Essig:2018pzq} to further narrow down parameter space of the inelastic SIDM model.

The paper is organized as the following. In Sec. 2, we present details of the model and outline the numerical method used in calculating elastic and inelastic dark matter self-scattering cross section. In Sec. 3, we present the astrophysical constraints on the parameter space. We conclude in Sec. 4. In the appendix, we provide details of substitutions and transformations exploited in this work for solving the Schr{\"o}dinger equation with two states.

\section{Particle Physics Model for inelastic SIDM}
\subsection{Scattering Cross Sections}

We assume that the dark matter particle is a fermion ($\Psi$) and it interacts with a dark U(1) gauge boson ($\phi_\mu$). The model can be described by the following Lagrangian~\cite{Finkbeiner:2007kk,ArkaniHamed:2008qn,Schutz:2014nka}
\begin{equation}
\begin{aligned}
\mathcal{L} = \bar{\Psi} (i\slashed{\partial} - m) \Psi - \frac{\Delta m}{4} ( \bar{\Psi} \Psi^{c} + \bar{\Psi}^{c} \Psi) + -\frac{1}{4} \phi^{\mu \nu} \phi_{\mu \nu} + \frac{1}{2} m_{\phi}^{2} \phi^{\mu} \phi_{\mu} + g_{\chi} \bar{\Psi} \gamma^{\mu} \Psi \phi_{\mu}  
\end{aligned}
\label{equ:lagrangian}
\end{equation}
where $\Psi^{c}$ is the charge conjugation of $\Psi$, $\phi_{\mu\nu}$ is the field strength of $\phi_\nu$, $g_\chi$ is the gauge coupling constant, $m$ is the Dirac mass of the dark matter state and $\Delta m$ is its Majorana mass. In this work, we assume $m\gg\Delta m$. Defining the Majorana mass eigenstates as $\chi_{1} = i(\Psi -  \Psi^{c})/\sqrt{2}$ and $\chi_{2} = (\Psi + \Psi^{c})/\sqrt{2}$, we rewrite equation (\ref{equ:lagrangian}) as  
\begin{equation}
\begin{aligned}
\mathcal{L} \supset \frac{1}{2} \bar{\chi}_{1} \left(i\slashed{\partial} - m_\chi\right) \chi_{1} + \frac{1}{2} \bar{\chi}_{2} \left(i\slashed{\partial} - (m_{\chi}+\Delta m)\right) \chi_{2} +\frac{i}{2} g_{\chi} \bar{\chi}_{2} \gamma^{\mu} \chi_{1} \phi_{\mu} + \rm{h.c.}
\end{aligned}
\label{equ:diagnolized_lagrangian}
\end{equation} 
where $m_{\chi} = m -\Delta m/2$ is the mass of the light state $\chi_{1}$.

\begin{figure}[t!]
\includegraphics[scale=0.3]{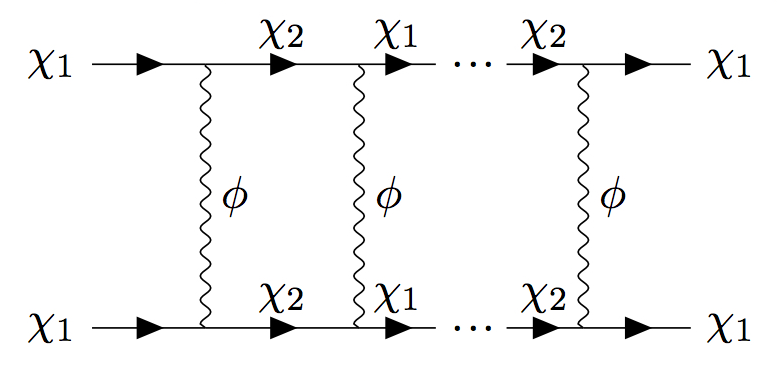}\;\;
\includegraphics[scale=1.26]{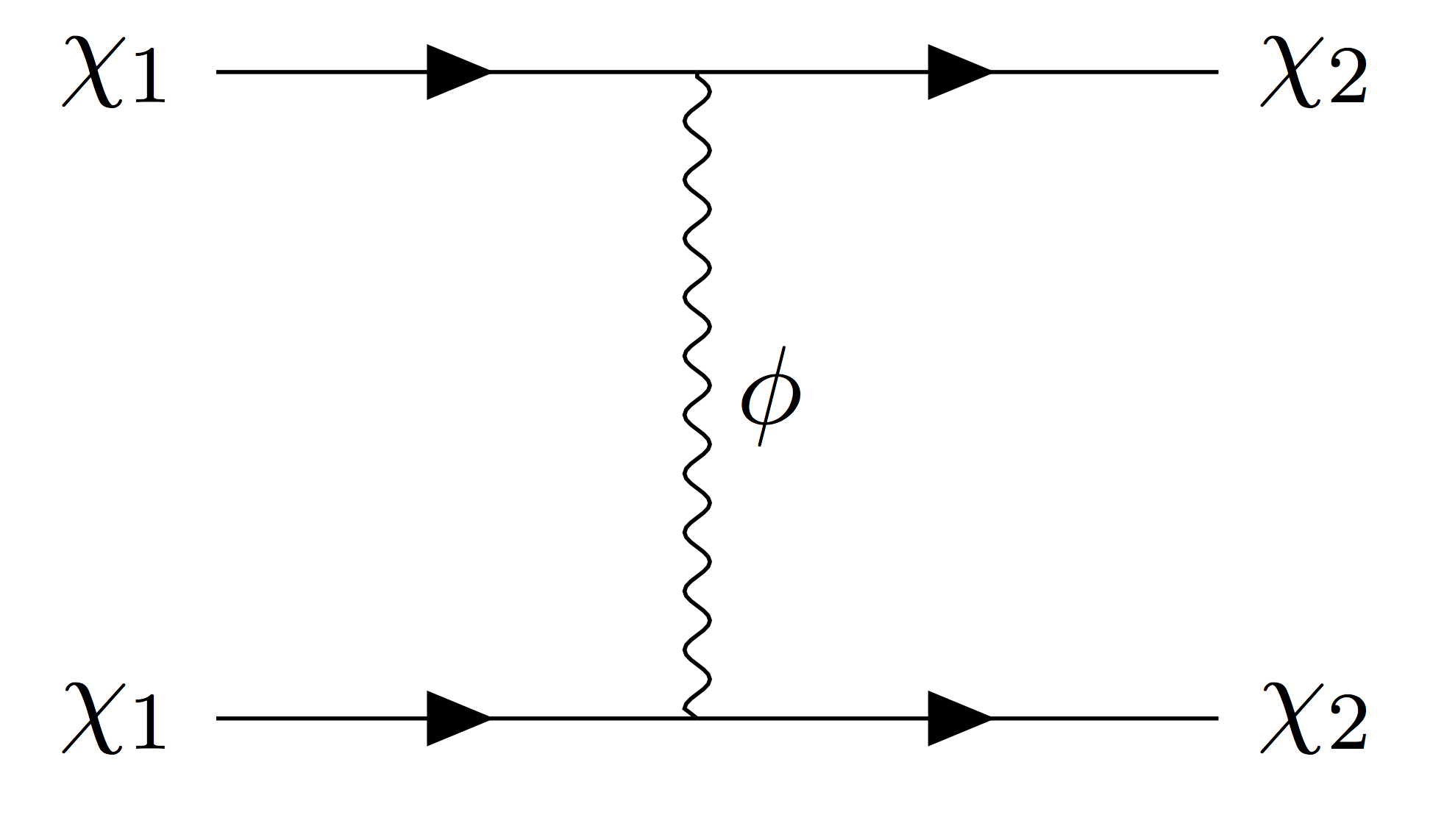}
\caption{ Feynman diagrams for elastic (left) and inelastic (right) dark matter self-interactions.}
\label{fig:feynman}
\end{figure}
Since the mass eigenstates are Majorana states, they carry no charge and only interact through an off-diagonal coupling. The relevant Feynman diagrams for both elastic and inelastic dark matter self-scattering are shown in Fig.~\ref{fig:feynman}. The tree-level elastic scattering process involves mixed initial and final states. All other elastic scatterings occur through high-order box or ladder diagrams. As the universe cools, the up-scattering process becomes kinematically unfavorable, driving the density of the heavy state down~\cite{Blennow:2016gde, Batell:2009vb}. Thus, we assume dark matter is made of the light state in the halo and only consider elastic and inelastic scattering processes shown in the left and right panels of Fig.~\ref{fig:feynman}, respectively. 

In the non-relativistic limit, we can apply the Schr{\"o}dinger formalism. There are two wave functions coupled by a matrix potential of the form
\begin{equation}
i \frac{\partial}{\partial t} \tilde{\psi} = \left[-\frac{1}{2 \mu} \nabla^{2}+\bold{V}\right]\tilde{\psi}
\label{equ:schrodinger}
\end{equation}
where $\mu= m_{\chi}/2$ is the reduced mass, the vector $\tilde{\psi}^{T} = \begin{bmatrix} \psi_{1} \ , \  \psi_{2} \end{bmatrix}$ detonate the wave functions for the two particle modes, and the matrix potential $\bold{V}$ is
\begin{equation}
\bold{V} =
         \begin{bmatrix}
           0 & -\frac{\alpha_{\chi}}{r} e^{- m_{\phi} r} \\
           -\frac{\alpha_{\chi}}{r}e^{- m_{\phi} r}& 2 \Delta m\\
         \end{bmatrix}.
         \label{equ:potential}
\end{equation} 
We have defined $\alpha_{\chi} \equiv g_{\chi}^{2} / 4\pi$ as the dark fine structure constant. The energy needed to create the heavy state as a pair is $2 \Delta m$. The numerical solution to this set of coupled differential equations gives the scattering cross sections through the method of partial waves. 

\subsection{Numerics}

We assume that dark matter freezes out in the early universe with the relic abundance to be consistent with the observed density. In this paper, we set the dark fine structure constant to $\alpha_{\chi} = 0.01\left(m_{\chi} /270~\rm{GeV}\right)$ such that the annihilation cross section is $6\times10^{-26}~{\rm cm^2/g}$. The dark matter self-scattering cross sections, both elastic and inelastic, are in general velocity-dependent. To capture the relevant physics on dwarf scales, we set the dark matter relative velocity to be $60~{\rm km/s}$ in the halo throughout this paper unless otherwise stated. The model is left with three free parameters, the dark matter mass $m_{\chi}$, the mass splitting $\Delta m$ and the mediator mass $m_{\phi}$.

Performing separation of variables on equation (\ref{equ:schrodinger}), we have the radial equation
\begin{equation}
\left[\frac{1}{r^{2}}\frac{\partial}{\partial r} \left(r^{2}\frac{\partial}{\partial r}\right)-\frac{l(l+1)}{r^{2}}+k^{2}\right] R_{l,i}(r) =m_{\chi} V_{i,j} R_{l,j}(r)
\label{equ:radial}
\end{equation}
where $l$ is the angular momentum mode, $k$ is the magnitude of the wave vector, $R_{l,i}(r)$ are the radial wave functions for $i = 1,2$ and $V_{i,j}$ denotes components of the matrix (\ref{equ:potential}).
Defining the following dimensionless parameters and substitutions, 
\begin{equation}
\begin{aligned}
x\equiv2 \alpha_{\chi} \mu r, \ \ 
a\equiv\frac{v}{2 \alpha_{\chi}}, \ \
b\equiv\frac{2 \alpha_{\chi} \mu}{m_{\phi}}, \ \
c^{2}\equiv a^{2}-\frac{\Delta m}{\mu \alpha_{\chi}^{2}}, \ \
\chi_{l,i}(x) \equiv x R_{l,i}(x)
\end{aligned}
\label{equ:parameters}
\end{equation}
we rewrite the radial equation (\ref{equ:radial}) in the matrix form as
\begin{equation}
\frac{d^{2}}{dx^{2}}
\begin{bmatrix}
\chi_{l,1} \\
\chi_{l,2} 
\end{bmatrix} 
= \begin{bmatrix}
\frac{l(l+1)}{x^{2}} - a^{2} & -\frac{1}{x} e^{-\frac{x}{b}} \\
-\frac{1}{x} e^{-\frac{x}{b}} & \frac{l(l+1)}{x^{2}} - c^{2}
\end{bmatrix} 
\begin{bmatrix}
\chi_{l,1} \\
\chi_{l,2} 
\end{bmatrix} .
\label{equ:radial_x}
\end{equation}
The wave function can be expanded in terms of spherical waves,
\begin{equation}
\tilde \psi=\sum_{l=0}^{\infty} (2l+1) P_{l}(\cos\theta) \left[\tilde \psi_{\rm{in}} \frac{e^{i p_{\rm{in}} x}-(-1)^{l}e^{-i p_{\rm{in}} x}}{2ip_{\rm{in}} x}+\begin{pmatrix} \alpha_{\chi} m_{\chi} \mathcal{F}_{x,l} \frac{e^{i a x}}{x} \\ \alpha_{\chi} m_{\chi} \mathcal{F}_{y,l} \frac{e^{i c x}}{x} \end{pmatrix}\right],
\label{equ:spherical_waves}
\end{equation}
where $\mathcal{F}_{x/y,l}$ are the scattering amplitudes for the two particle system and $p_{\rm{in}}=a,c$, depending on the initial state. If the initial state is $\chi_1$ as we consider in this work, $p_{\rm{in}}=a$ and $\tilde \psi_{\rm{in}}^{T}=\begin{bmatrix} 1 \ , \ 0 \end{bmatrix}$. The differential cross section is given by,
\begin{equation}
\frac{d \sigma}{d \Omega} = \frac{p_{\rm{out}}}{p_{\rm{in}}} \left| \sum_{l=0}^{\infty} (2l+1) P_{l}(\rm{cos}\theta)\mathcal {F}_{\it l}\right|^{2}
\label{equ:diff_xsection}
\end{equation}
where $p_{\rm{in}}$, $p_{\rm{out}}$ and $\mathcal{F}_{l}$ again depend on the initial and final states.
For elastic scattering $\chi_{1} \chi_{1} \rightarrow \chi_{1} \chi_{1}$, $p_{\rm{in}}=p_{\rm{out}} = a$ and $\mathcal{F}_{l} =\mathcal{F}_{x,l}$.
For inelastic scattering $\chi_{1} \chi_{1} \rightarrow \chi_{2} \chi_{2}$, $p_{\rm{in}}=a, \ \ p_{\rm{out}} = c$ and $\mathcal{F}_{l} =\mathcal{F}_{y,l}$.

\begin{figure}[tbp]
   \includegraphics[scale=0.35]{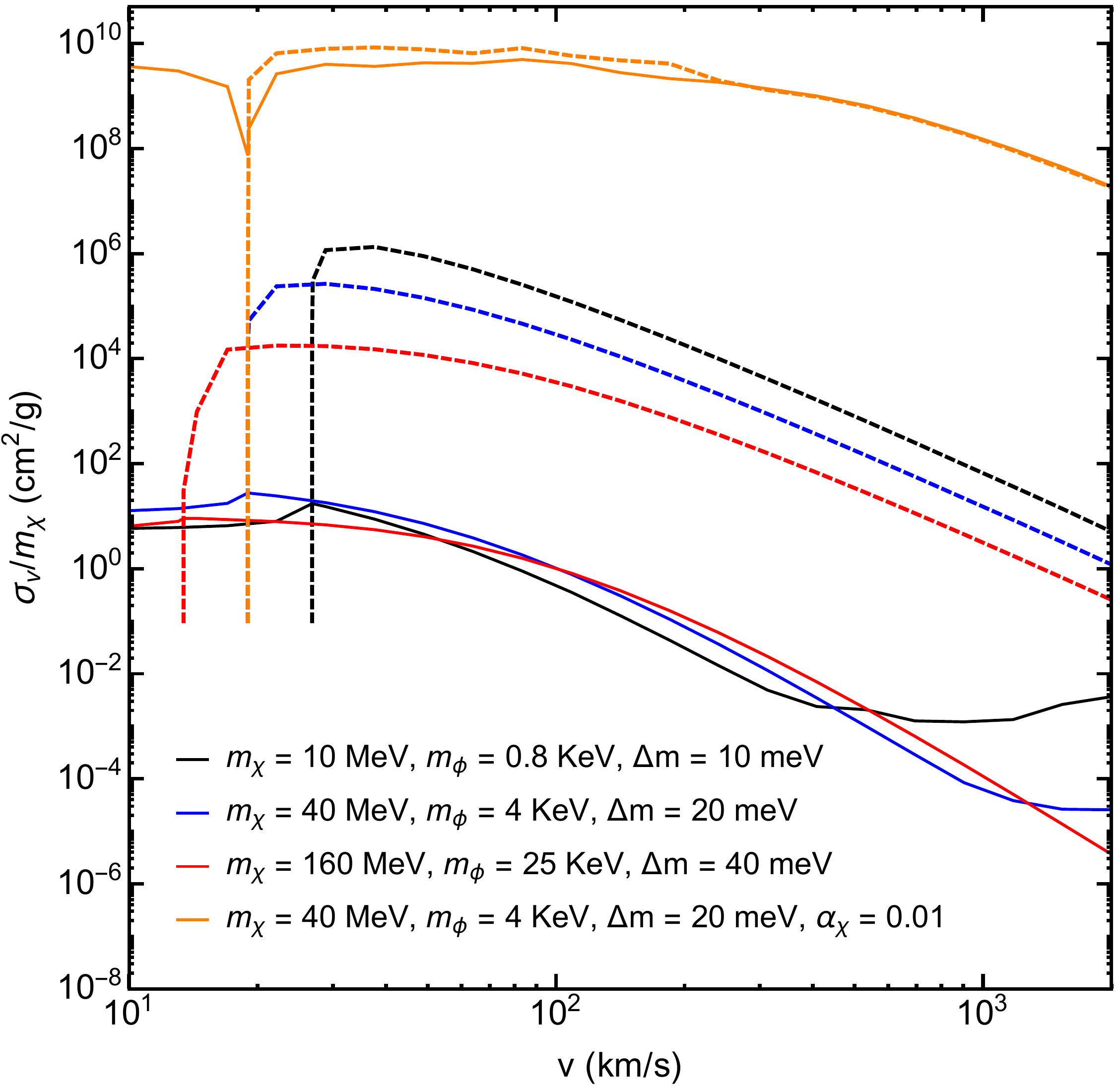}\;\;
   \includegraphics[scale=0.35]{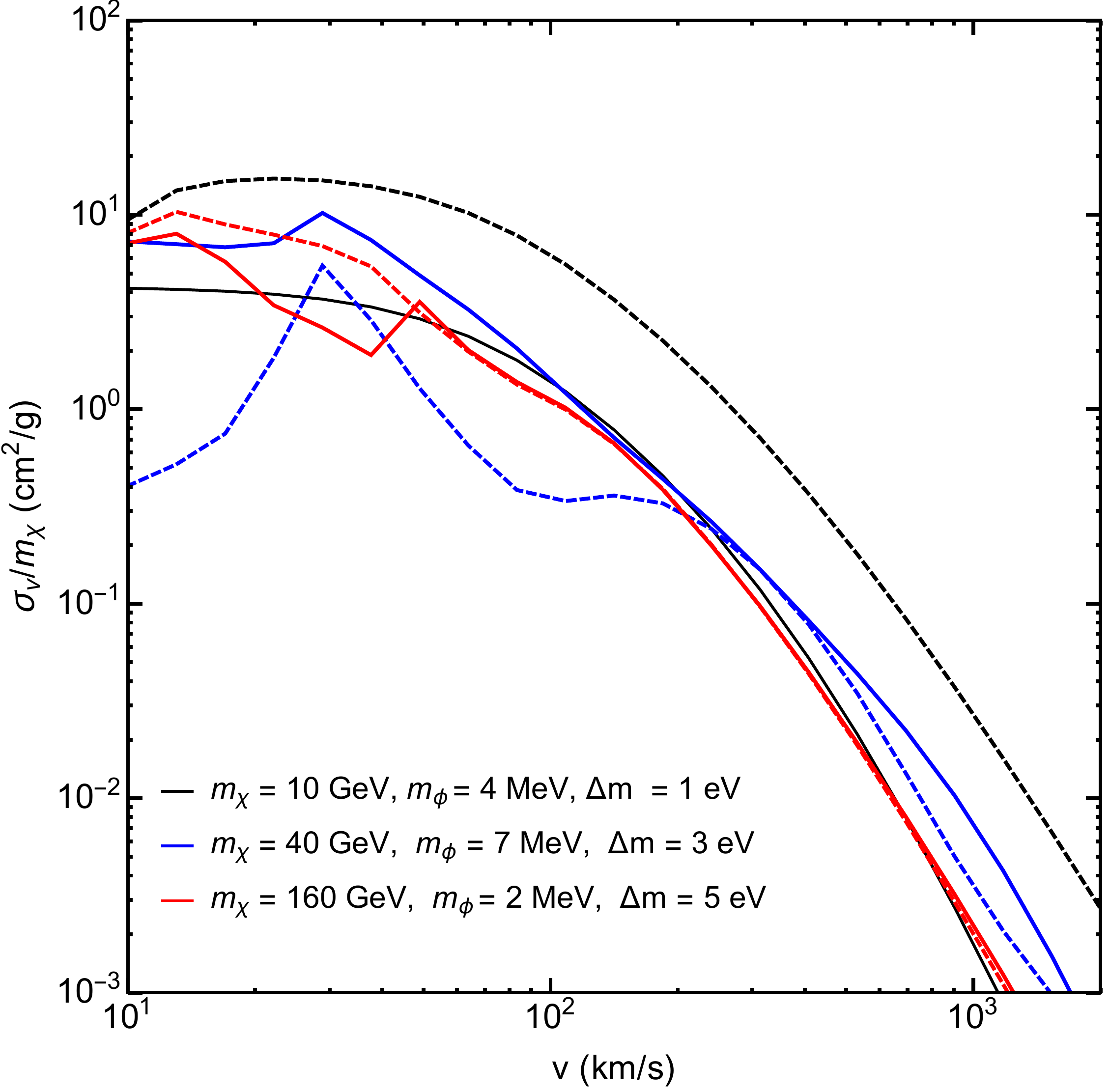}
\caption{The dark matter self-scattering cross section vs  the relative scattering velocity for benchmark cases with the dark matter mass in the ${\rm MeV}$ (left panel) and ${\rm GeV}$ (right panel) ranges, where we fix the dark coupling constant using the relic abundance relation, $\alpha_{\chi} = 0.01(m_{\chi} /  270 \ \rm{GeV})$. The solid and dotted curves correspond to the elastic and inelastic cross sections, respectively. For comparison, we also show a case for $m_\chi=40~{\rm MeV}$ and $\alpha_{\chi}=0.01$ in the left panel.}
\label{fig:velocity}
\end{figure}

To find the scattering amplitudes $\mathcal {F}_{x/y, l}$, we need to first find the wave function by numerically solving equation (\ref{equ:radial_x}), and then map its form at large radii onto the spherical wave expansion in equation (\ref{equ:spherical_waves}). However, a direct numerical solution to the wave equation (\ref{equ:radial_x}) is unstable for a large part of the parameter space of interest. To tame these instabilities, we follow the procedure discussed in~\cite{Ershov:2011zz,Blennow:2016gde} and make a number of substitutions to transform the wave equation into a more manageable form; see the appendix for details. In this work, we calculate the viscosity cross section for dark matter self-interactions~\cite{Tulin:2013teo},
\begin{equation}
\sigma_{V} = \int{d\Omega \frac{d\sigma}{d\Omega} \rm{sin}^{2}\theta},
\end{equation}
which regulates both forward and backward scatterings. See the appendix for an explicit expression of the viscosity cross section in terms of phase shifts.

In Fig.~\ref{fig:velocity}, we show the elastic (solid) and inelastic (dashed) self-scattering cross sections vs the relative velocity for a few representative cases, where we choose a wide range of dark matter masses from $10~{\rm MeV}$ to $160~{\rm GeV}$. Overall the cross sections decrease as the velocity increases. For dark matter masses below $1~{\rm GeV}$ (left panel), there is a clear indication of the threshold velocity below which inelastic up scattering is kinematically forbidden. In this mass range, when the coupling constant is set by the relic abundance constraint~\cite{Boehm:2003hm,Pospelov:2007mp,Feng:2008ya,Feng:2008mu}, i.e., $\alpha_{\chi} = 0.01(m_{\chi} /  270 \ \rm{GeV})$, the inelastic scattering cross section is much larger than the elastic one as long as the up-scattering channel is open. As we discussed, in this model, inelastic up scattering occurs at the tree-level, while elastic scattering at the high-order level. For small $m_\chi$ below $1~{\rm GeV}$, the dark fine structure constant $\alpha_{\chi}$ is small as well, and the non-perturbative quantum effect is absent to enhance the elastic cross section. To demonstrate this point, we present another case, where we set $\alpha_{\chi}=0.01$ for $m_\chi=40~{\rm MeV}$. In this case, both elastic and inelastic cross sections are similar for the velocity larger than $18~{\rm km/s}$. For high dark matter masses (right panel), the elastic and inelastic cross sections become more compatible, aside from resonance peaks, and the up-scattering process is kinematically allowed in the plotted velocity range. As $m_\chi$ increases, the gauge coupling $\alpha_\chi$ increase accordingly and the non-perturbative effect boosts the elastic scattering cross section significantly.

\section{Astrophysical Constraints}

\begin{figure}[t!]
\includegraphics[scale=0.4]{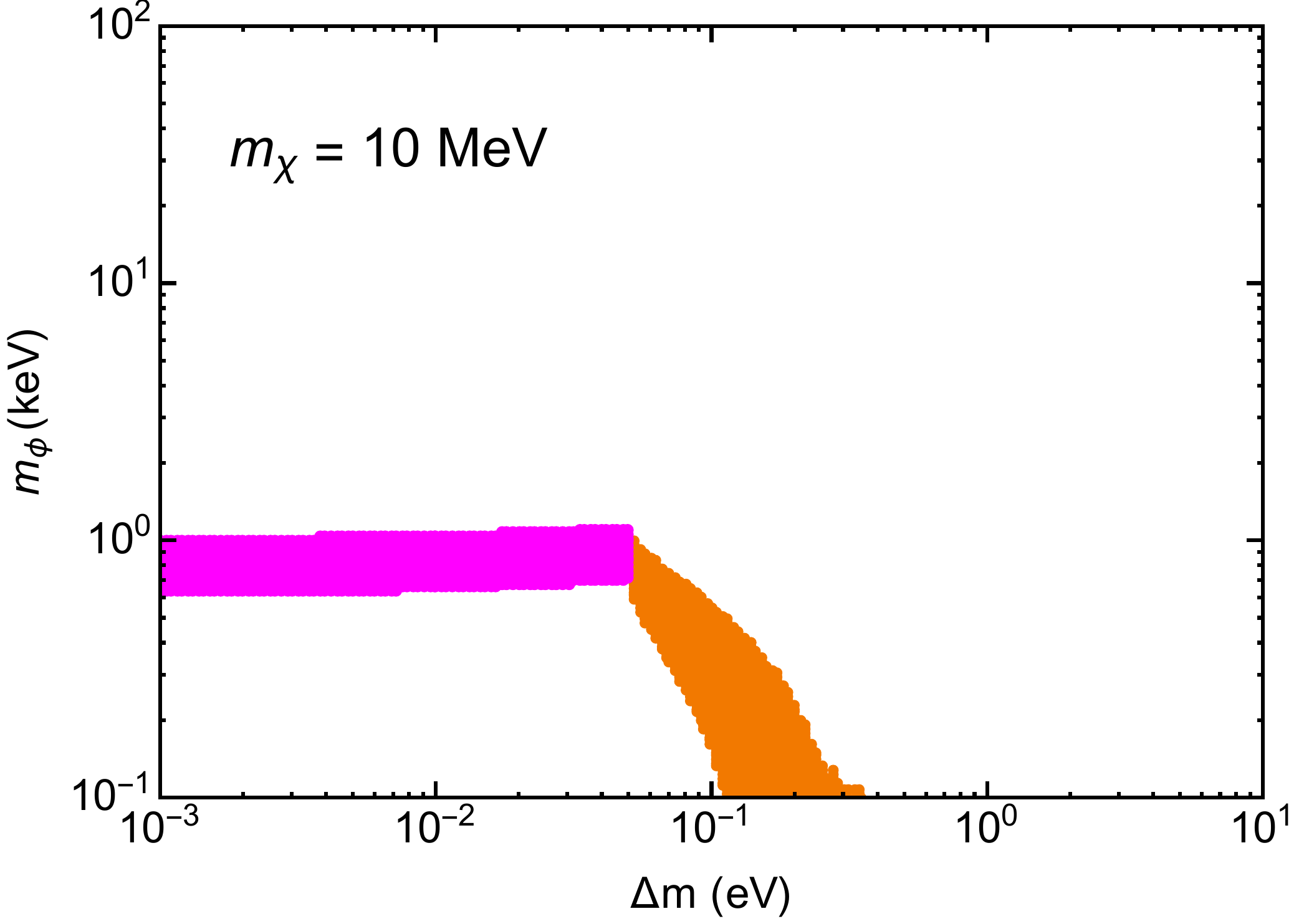}
\includegraphics[scale=0.4]{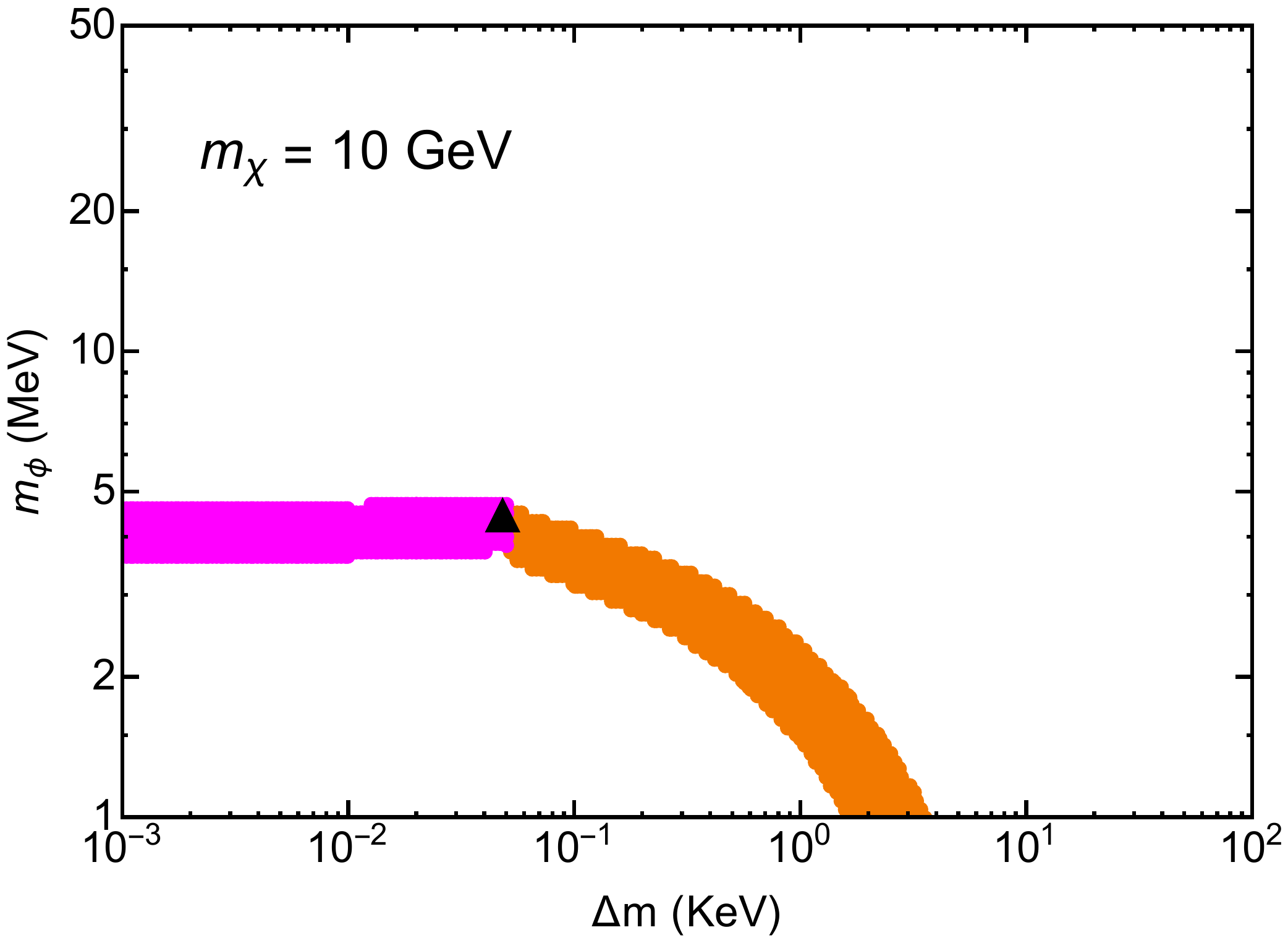}
\includegraphics[scale=0.4]{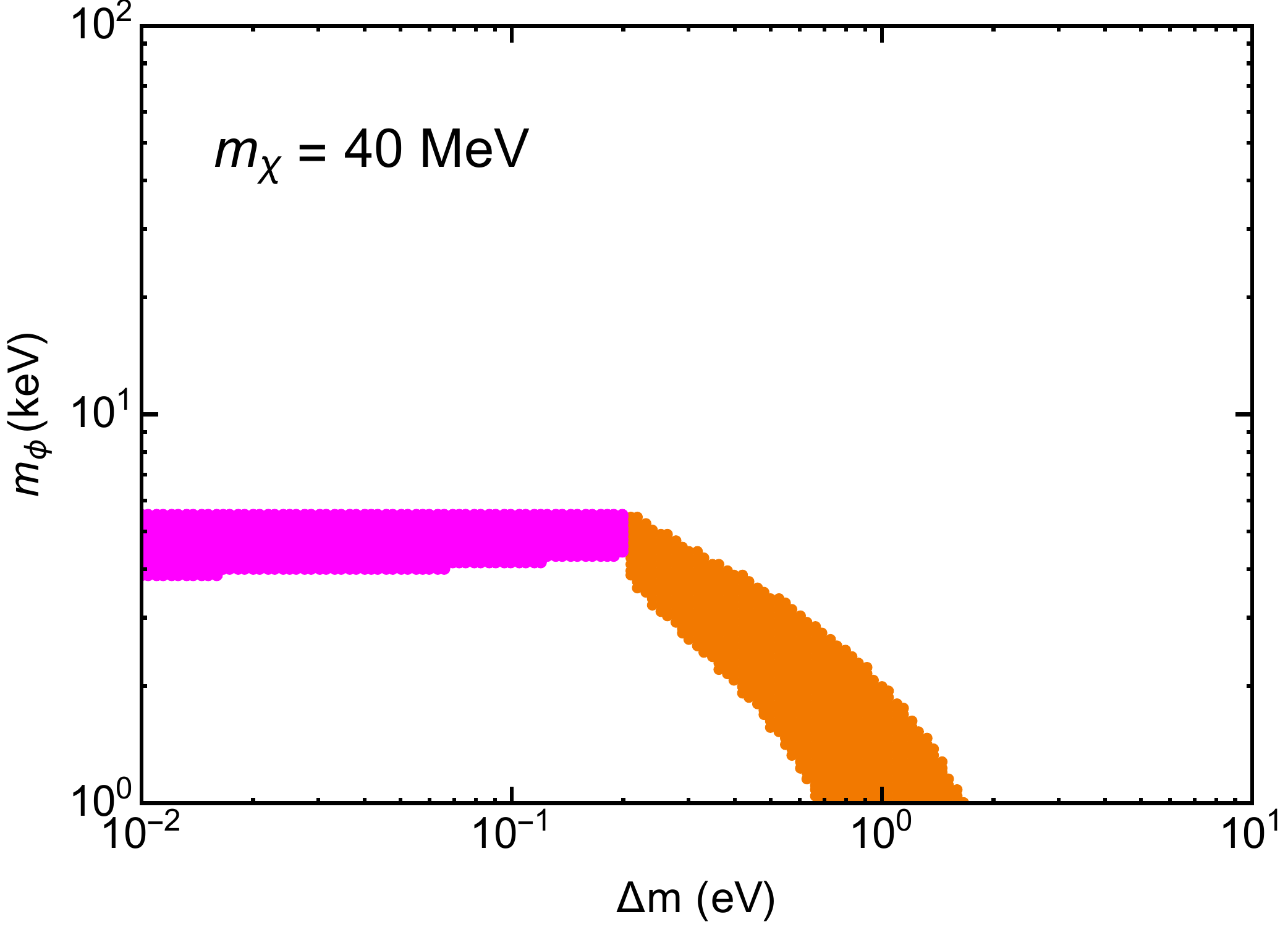}
\includegraphics[scale=0.4]{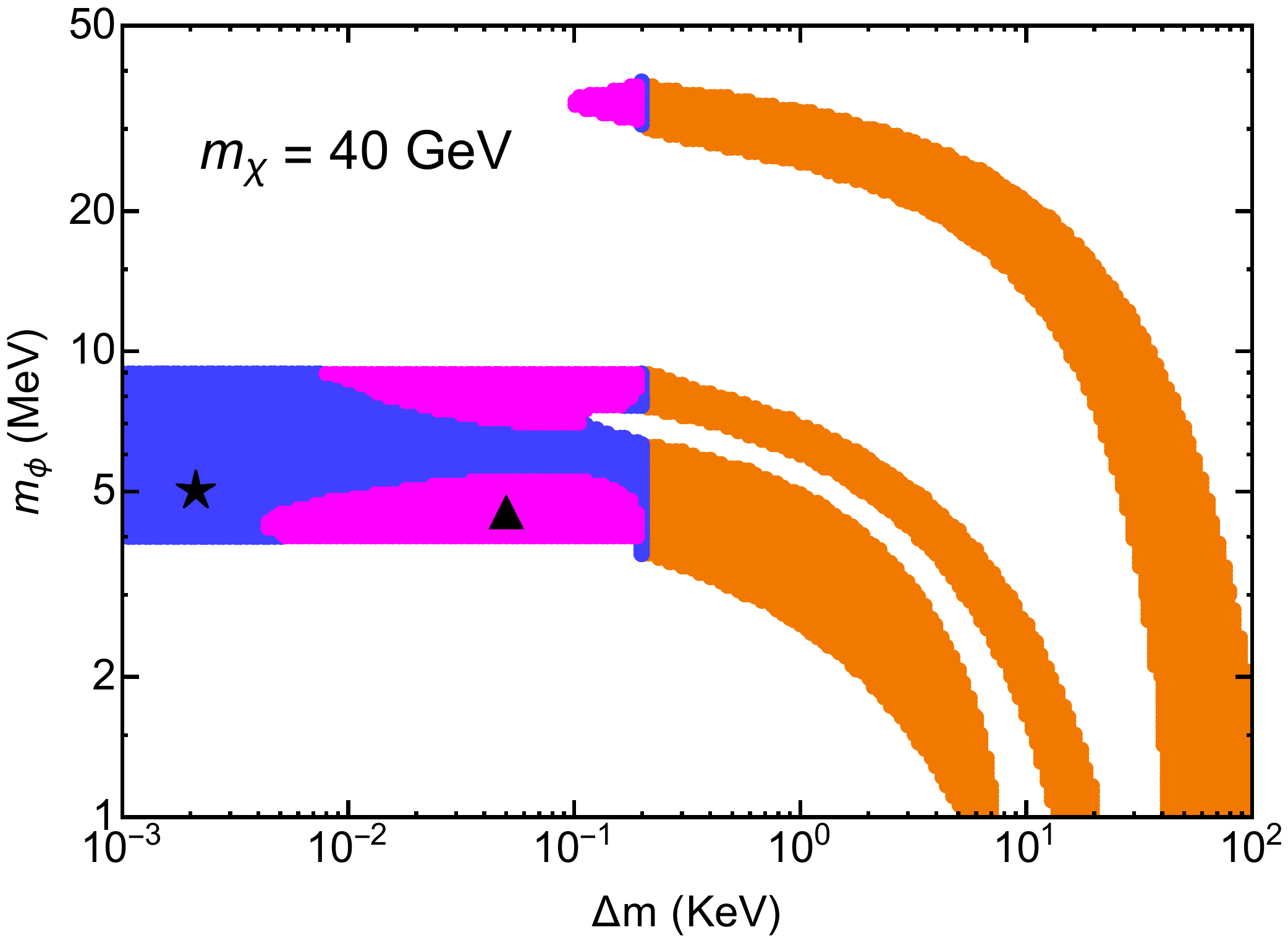}
\includegraphics[scale=0.4]{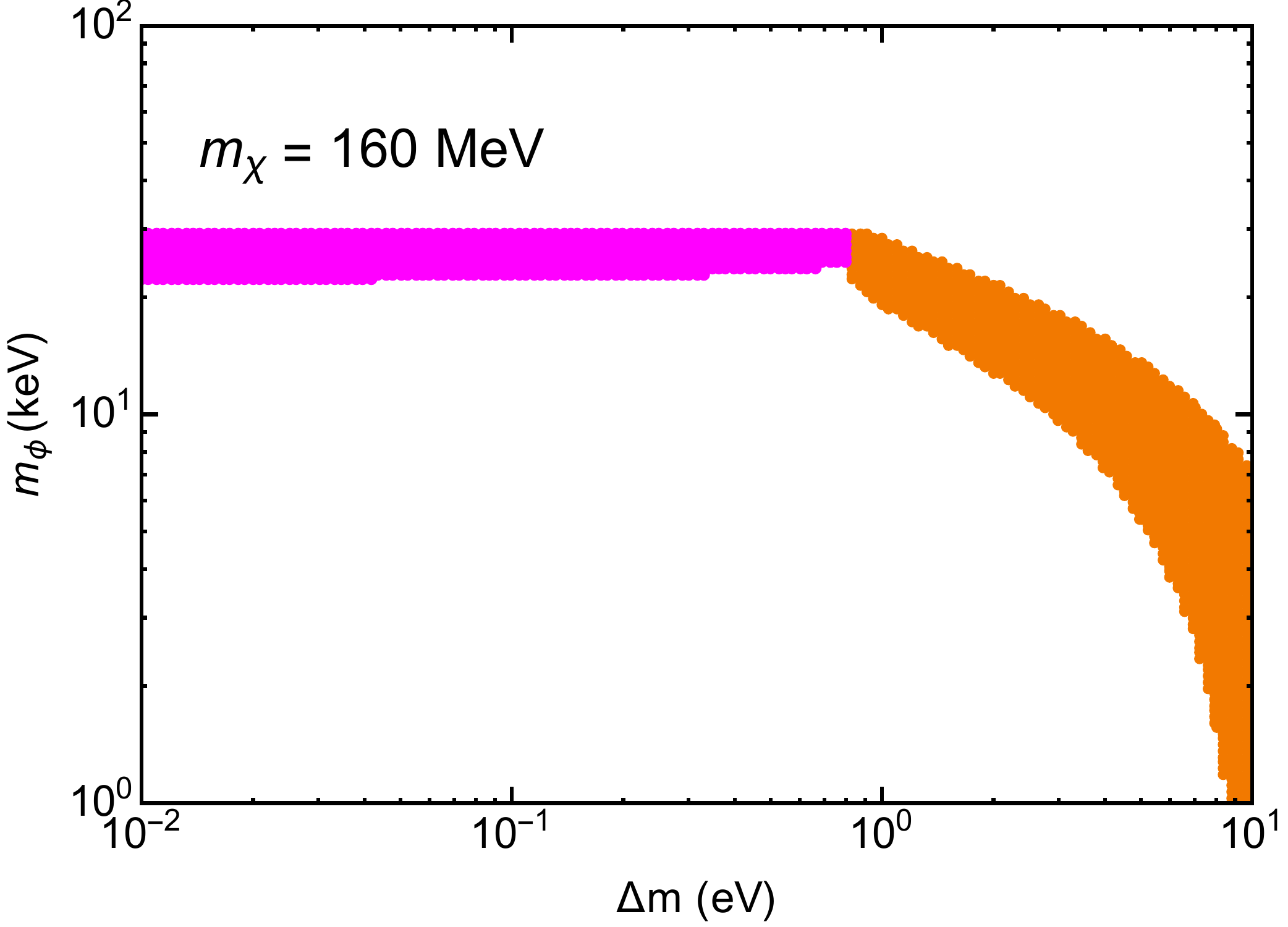}
\includegraphics[scale=0.4]{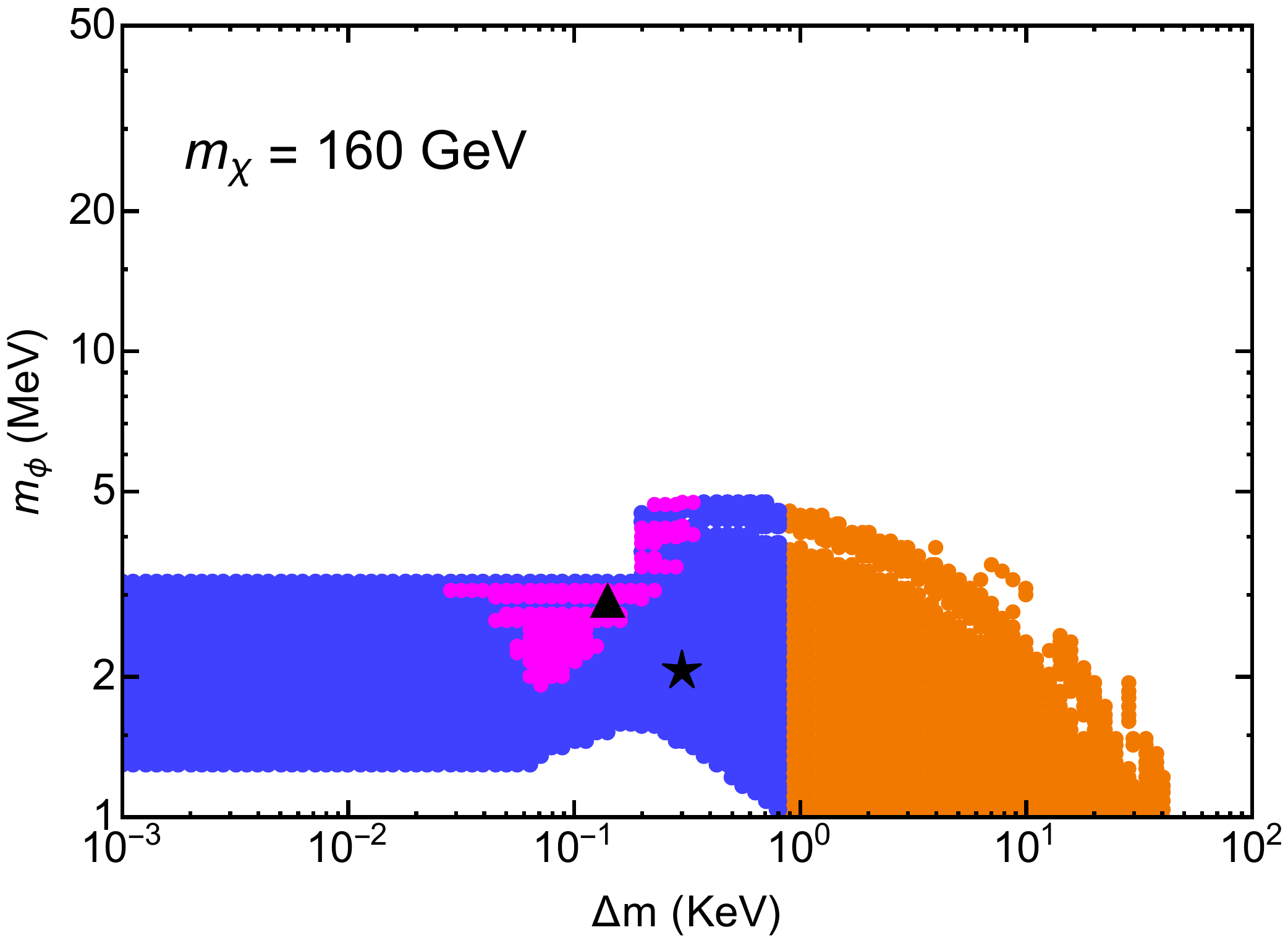}
\caption{The shaded regions correspond to the accessible parameter space in the $\Delta m\textup{--}m_\phi$ plane, where the elastic scattering cross section is in the range of $1\leq\sigma_V / m_{\chi}\leq5~\rm{cm^2/g}$, favored by solving the small-scale issues. We have set the dark matter relative velocity to be $60~{\rm km/s}$, a characteristic value in dwarf galaxies that prefer a dark matter density core. In the orange regions, the up-scattering process ($\chi_1\chi_1\rightarrow\chi_2\chi_2$) is kinematically forbidden and dark matter self-interactions are purely elastic. In the magenta (blue) regions, the up-scattering process is allowed, and the dissipation process associated with the $\chi_2$ decay can lead to core collapse in dwarf galaxies with a timescale shorter (longer) than $10~{\rm Gyr}$. The starred and triangle points are references which show the mapping between the elastic scattering and the core-collapse constraints, as explicitly shown in Fig.~\ref{fig:constraints}. }
\label{fig:parspace}
\end{figure}

Taking the benchmark $m_\chi$ values shown in Fig.~\ref{fig:velocity}, we scan parameter space of the $\Delta m\textup{--}m_\phi$ plane, such that the elastic cross section fall within the range of $1~{\rm cm^2/g}\leq \sigma_V/m_{\chi} \leq 5~\rm{cm^{2}/g}$ for the relative velocity $60~{\rm km/s}$, a characteristic value for dwarf galaxies that prefer a dark matter density core. In Fig.~\ref{fig:parspace}, we show the resulting parameter space (shaded). For a given dark matter mass, there is a preferred range in the plane, where the elastic self-scattering cross section is large enough to thermalize the inner halo in accord with observations~\cite{Tulin:2017ara}. For all cases, if the mass splitting $\Delta m$ is small, the mediator mass $m_\phi$ is almost a constant. While, as $\Delta m$ increases towards the high end, $m_\phi$ must decrease to preserve the elastic cross section in the desired range, since the elastic process involves virtual up-scattering processes, as shown in Fig.~\ref{fig:feynman} (left). The transition occurs when the mass splitting reaches the kinematic threshold, where up scattering is forbidden for larger values of $\Delta m$, i.e., $2 \Delta m = \mu v^{2}_{\rm rel}/2$ with $v_{\rm rel}=60~{\rm km/s}$. In Fig.~\ref{fig:parspace}, the orange shaded regions are where $\chi_1\chi_1\rightarrow\chi_2\chi_2$ is kinematically forbidden, while in the magenta and blue regions, the up scattering is allowed. Note in the  case of $m_\chi=40~{\rm GeV}$ there is more than one branch for the favored parameter space, because the scattering is in the strong resonance regime~\cite{Tulin:2012wi,Tulin:2013teo} and multiple ranges of the mediator mass is allowed; see also~\cite{Blennow:2016gde}.

If the mass splitting is large enough and up scattering is forbidden, dark matter self-interactions are purely elastic and the condition of $1~{\rm cm^2/g}\leq\sigma_{V}/m_\chi\leq5~{\rm cm^2/g}$ is sufficient enough to specify astrophysical constraints. However, if $\chi_1\chi_1\rightarrow\chi_2\chi_2$ is allowed in the halo and the resulting $\chi_2$ can further decay to $\chi_1$ and some light species, this dissipative process may cool the inner halo and speed up the SIDM core collapse~\cite{Essig:2018pzq}. For the model we consider, $m_\phi\gg\Delta m$ in the parameter regions of interest, e.g., the shaded regions in Fig.~\ref{fig:parspace}, hence the decay process $\chi_2\rightarrow\chi_1\phi$ is kinematically forbidden. On the other hand, if we consider a more general setup, there are other interaction terms that may lead to dissipative decays of $\chi_2$. For example, Ref.~\cite{Finkbeiner:2014sja} introduces a dimension-$5$ dipole operator $(1/M)\bar{\chi_{2}}\sigma^{\mu\nu}{\chi_{1}}F_{\mu\nu}$, where $M$ is the cut-off scale and $F_{\mu\nu}$ is the field strength of the standard model photon. With this operator, $\chi_2$ can decay to $\chi_1$ and $\gamma$. The rate is $\Gamma_{\chi_{2}\rightarrow\chi_{1}\gamma} =  4\Delta m^{3}/(\pi M^{2})$, and $\chi_2$'s lifetime is $\tau = 1/\Gamma_{\chi_{2}\rightarrow\chi_{1}\gamma}  \sim 0.5~\rm{sec}\left( {\it M} / {\rm TeV} \right)^{2}\left(\rm{keV} / \Delta m \right)^{3}$. For $\Delta m \sim 10^{-3} \rm{eV}$, it is comparable to the age of galaxies, $\sim10~{\rm Gyr}$, for $M$ up to $\sim 1~\rm{TeV}$. Thus, this dissipative decay is relevant to halo dynamics if the dipole operator is present. In addition, in atomic dark matter models, an excited atomic state can decay to a ground state by emitting a massless dark photon. 

\begin{figure}[t!]
\includegraphics[scale=0.33]{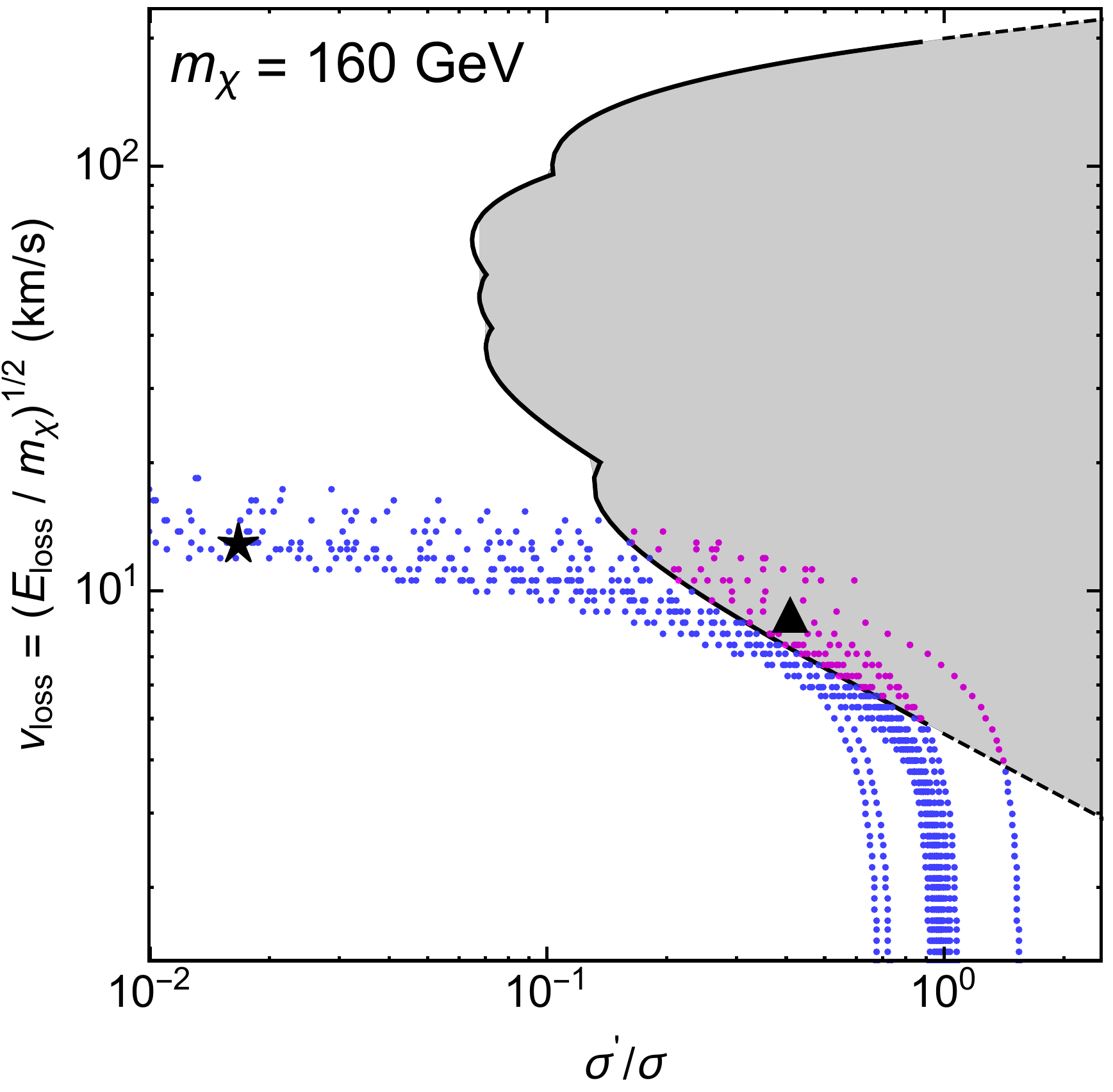}\;
\includegraphics[scale=0.33]{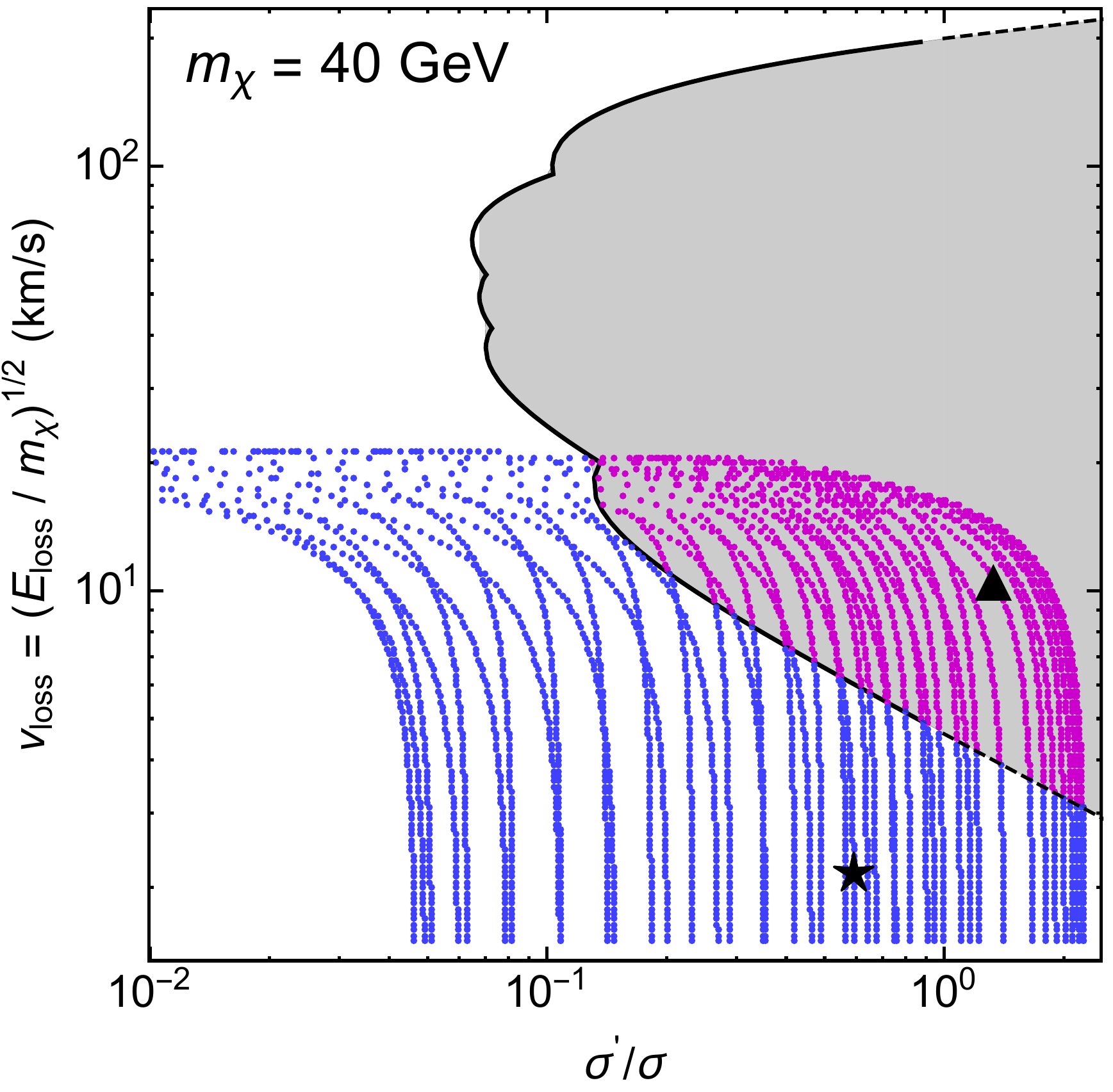}\;
\includegraphics[scale=0.33]{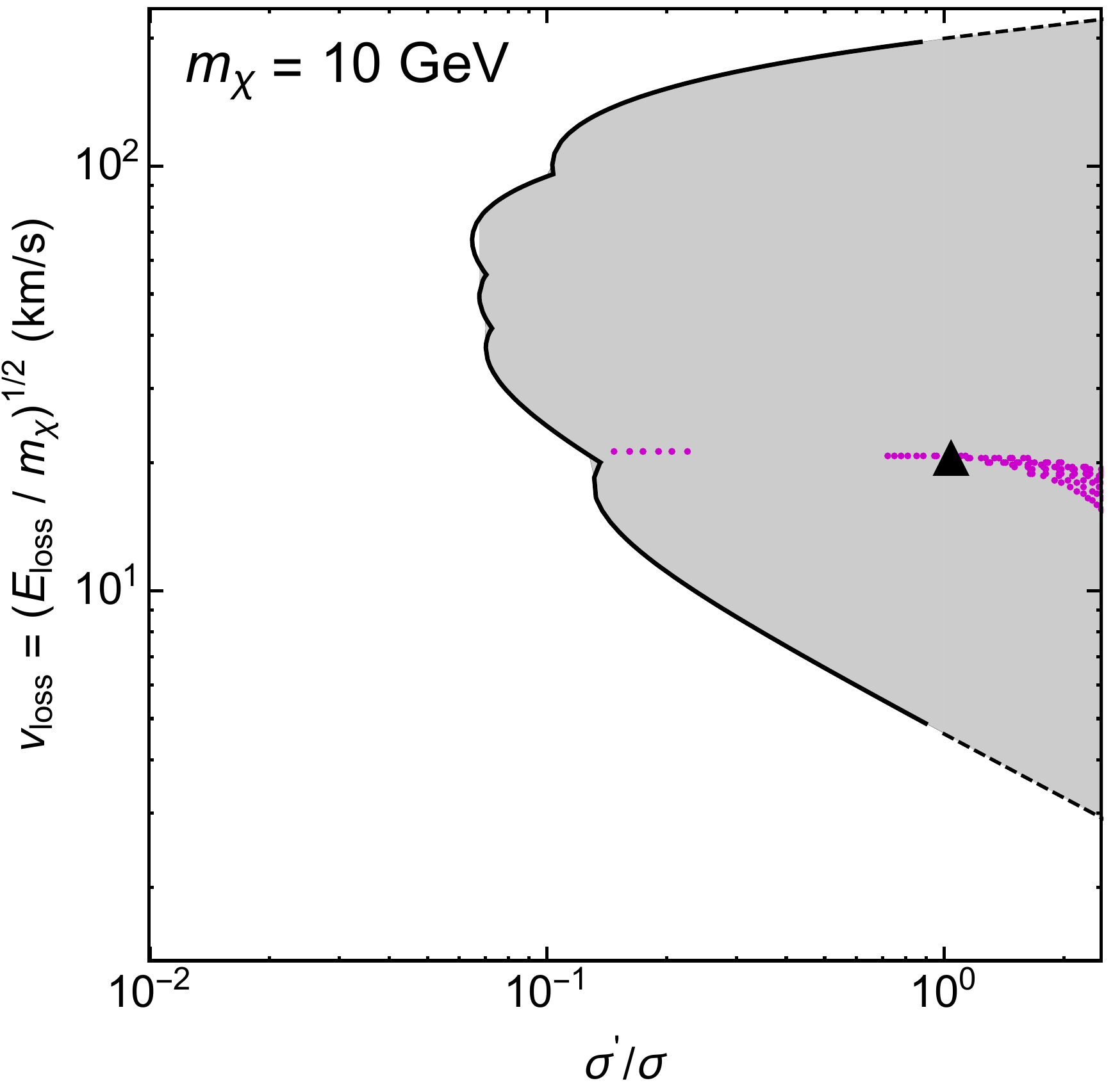}
\caption{Mapping between elastic scattering and galaxy core-collapse constraints on the $ \sigma ' / \sigma \textup{--}\nu_{\rm{loss}}$ plane. In the gray regions within the black contour, the halo core collapse induced by dissipative dark matter self-interactions occurs within the age of galaxies; adapted from \cite{Essig:2018pzq}. The dashed black lines are the linear extrapolation of the contour for $\sigma ' / \sigma >1$. The magenta points that lie within the gray regions are disfavored by the core-collapse constraints; while the blue points outside are still allowed. The stars and triangles are reference points and their correspondences are shown on the $\Delta m\textup{--}m_\phi$ plane in Fig.~\ref{fig:parspace}}. 
\label{fig:constraints}
\end{figure}

In what follows, we assume that $\chi_2$ can decay to $\chi_1$ and a massless species that escapes the halo, and study additional astrophysical constraints on the parameter space. Using dwarf galaxies that show shallow density cores, Ref.~\cite{Essig:2018pzq} derives bounds on dissipative dark matter interactions by demanding the core-collapse timescale longer than the age of galaxies, $\sim10~{\rm Gyr}$. In particular, it uses the energy loss per collision and the ratio of inelastic to elastic cross sections, i.e., $\nu_{\rm{loss}} = \sqrt{E_{\rm{loss}} / m_{\chi}}$ and $\sigma'/\sigma$, respectively, to characterize the cooling effect, and places constraints on their combinations. To apply the core-collapse constraints on our model, we set $E_{\rm{loss}} = \Delta m$, calculate $\nu_{\rm loss}$ and $\sigma'/\sigma$ values for each favored model point shown in Fig.~\ref{fig:parspace} (shaded), and then compare them with the limits on the $\sigma'/\sigma\textup{--}\nu_{\rm loss}$ plane from~\cite{Essig:2018pzq} as reproduced in Fig.~\ref{fig:constraints} (gray shaded). In the magenta shaded regions of Fig.~\ref{fig:parspace}, the dissipative self-interactions are strong enough to cause core collapse in dwarf halos within $10~{\rm Gyr}$. While in the blue regions, inelastic up scattering can occur, but the overall cooling rate is small to trigger core collapse in the age of galaxies. 

To better understand these constraints, we show the distribution of the model points in the $\sigma'/\sigma\textup{--}\nu_{\rm loss}$ plane for three benchmark cases in Fig.~\ref{fig:constraints}, along with the bounds from~\cite{Essig:2018pzq} (gray). All points (magenta) that lie within the gray regions are disfavored as they result in a core-collapse timescale too short to fit the observations; while the points (blue) outside are still allowed. We classify the model points shown in Fig.~\ref{fig:parspace} using the same color scheme. Note we have extrapolated the disfavored parameter space following the trend beyond the upper limit of $\sigma'/\sigma$ in~\cite{Essig:2018pzq}  (black dashed). This is reasonable, because the bounds should be stronger as $\sigma'$ further increases. 

From Fig.~\ref{fig:constraints}, we see that as the dark matter decreases from $160~{\rm GeV}$, more of the parameter space is disfavored by the core-collapse constraints. When the mass approaches $10~{\rm GeV}$ or smaller, the entire model points lie within the gray regions. Since the gauge coupling reduces as the dark matter mass decreases, the inelastic scattering gradually dominates over the elastic one if the former is open. For the dark matter below $\sim10~{\rm GeV}$, only the portion of the parameter space, where inelastic scattering is kinematically forbidden, remains viable. While for the cases of $m_\chi=160~{\rm GeV}$ and $40~{\rm GeV}$, in some parts of the parameter space, inelastic up scattering is allowed, but the cooling rate is not significant so they evade the collapse constraints, because either the inelastic cross section or the energy loss per collision is small. We demonstrate this by using the reference points in both Figs.~\ref{fig:parspace} and~\ref{fig:constraints} (stars and triangles), which are in one-to-one correspondence. In the case of $m_\chi=160~{\rm GeV}$, the two references have similar $\Delta m$ and $\sigma/m$, but the star point has much smaller $\sigma'/m$ than the triangle one as the former is closer to the threshold of the up scattering. While in the case of $m_\chi=40~{\rm GeV}$, the reference points mainly differ in $\Delta m$, resulting different locations in the $\sigma'/\sigma\textup{--}\nu_{\rm loss}$ plane.

\section{Conclusion}

We have studied an inelastic dark matter model with a light mediator, based on a U(1) gauge symmetry. The presence of a small Majorana mass splits a Dirac fermion into two Majorana states and the light one is the dark matter candidate. In this model, both elastic and inelastic dark matter self-interactions can be present in the halo. The former is mediated by a high-order process with multiple exchanges of the light mediator; while the latter by a tree-level process when it is kinematically allowed. Using the technique of partial waves, we have developed a numerical procedure to calculate the elastic and inelastic scattering cross sections.

We have explored astrophysical constraints on the model parameters, i.e., the dark gauge coupling constant, dark matter mass, mediator mass and mass splitting between the two Majorana states. We first imposed the relic density constraint on the coupling constant by assuming the standard freeze-out scenario, then chose six benchmark cases that cover a wide range of the dark matter mass, $10~{\rm MeV}\textup{--}160~{\rm GeV}$. For each case, we have found the parameter regions, where the elastic scattering cross section falls within the range of $1~{\rm cm^2/g}\textup{--}5~{\rm cm^2/g}$ in dwarf galaxies in order to solve the small-scale issues. Our analysis shows that if the mass splitting gets too large, the kinematic suppression of the intermediate virtual processes demand that the mediator become lighter to preserve the desired elastic cross section. We also found that when the dark matter mass decreases the inelastic scattering cross section dominates over the elastic one. This is because the coupling constant becomes smaller as the mass decreases and the non-perturbative quantum enhancement for the elastic cross section diminishes accordingly.  

If the heavy state can decay to the light state and a massless degree of freedom, inelastic dark matter self-interactions may induce a dissipative process that cools the inner halo and leads to SIDM core collapse. Observations of dark matter density cores in many low surface brightness galaxies put a constraint on the rate of energy loss. We studied its implications for the dark matter model we consider, and found it eliminates the majority of the parameter space for dark matter masses below $\sim{10}~{\rm GeV}$, unless the mass splitting is large enough so that the up-scattering process is forbidden. For a higher mass, there are parameter regions where the model evades the core-collapse constraints while the inelastic scattering is kinematically allowed. Our work demonstrates that astrophysical observations can provide powerful tests for inelastic dark matter models with a light mediator. The analysis can be used to constrain models where dark matter is made of a composite state, such as dark atoms and nuclei. It is also interesting to test those models using observations of galaxy clusters that show evidence of a density core in their inner halos~\cite{Newman:2012nw,Kaplinghat:2015aga}.

\appendix
\section{Cross Section Calculations}
\subsection{Reformulation}

In this appendix, we outline the variable phase space approach to reformulate equation (\ref{equ:radial_x}); see~\cite{Ershov:2011zz} for a more detailed discussion on this method. The basic idea is to build a solution to equation (\ref{equ:radial_x}) using solutions to the free-particle case ($\alpha_{\chi} \rightarrow 0$). In the non-interacting limit, the solutions can be written as superpositions of spherical Bessel and Neumann functions with constant coefficients. To build solutions to (\ref{equ:radial_x}), we use superpositions of free-particle solutions and upgrade the coefficients to functions, i.e.,
\begin{equation}
\chi^{(l)}_{i}(x) = \alpha^{(l)}_{i}(x) f^{(l)}(p_{i} x) - \beta^{(l)}_{i}(x) g^{(l)}(p_{i}x),
\label{equ:guess}
\end{equation} 
where $\chi^{(l)}_{i}(x)$ are the component solutions to (\ref{equ:radial_x}), $\alpha^{(l)}_{i}(x)$ and $\beta^{(l)}_{i}(x)$ are numerical functions, $p_{i} =a,c$, depending on the particle state, and $f^{(l)}(p_{i}x)$ and $g^{(l)}(p_{i}x)$ are the free-particle solutions. They obey the differential equation,
\begin{equation}
\left[\frac{d^{2}}{dx^{2}}-\frac{l(l+1)}{x^{2}}+p_{i}^{2}\right] z^{(l)}(p_{i}x) = 0.
\label{equ:free_solution}
\end{equation}
where $z$ takes the place of $f$ or $g$. 
The function $f^{(l)}(p_{i}x)$ is defined to be regular at the origin and $g^{(l)}(p_{i}x)$ is irregular as $x \rightarrow 0$.

To form a general solution to equation (\ref{equ:radial_x}), we must solve two coupled second-order differential equations. We therefore require four linearly independent solutions. The expression (\ref{equ:guess}) represents only one of the four solutions but has four degrees of freedom; two of them come from $\alpha^{(l)}_{i}(x)$ and $\beta^{(l)}_{i}(x)$ and the other two from the normalization of $ f^{(l)}(p_{i}x)$ and $ g^{(l)}(p_{i}x)$. We must impose constraints to reduce the extra degrees of freedom. Suppose that $d\chi^{(l)}_{i}(x)/dx$ is independent of the derivatives of $\alpha_{i}^{(l)}(x)$ and $\beta_{i}^{(l)}(x)$, which is trivially true for constant coefficients.
This requires that
\begin{equation}
\frac{d\alpha_{i}^{(l)}(x)}{dx} f^{(l)}(p_{i}x)-\frac{d\beta_{i}^{(l)}(x)}{dx} g^{(l)}(p_{i}x)=0.
\label{equ:constraint_alpha_beta}
\end{equation}

We set the normalization of $f^{(l)}(p_{i}x)$ and $g^{(l)}(p_{i}x)$ by defining the Wronskian of the system to be 
\begin{equation}
\frac{df^{(l)}(p_{i}x)}{d(p_{i}x)}g^{(l)}(p_{i}x)-f^{(l)}(p_{i}x)\frac{dg^{(l)}(p_{i}x)}{d(p_{i}x)} \equiv p_{i}.
\label{equ:normalization}
\end{equation}
After imposing the constraints, we have only one degree of freedom and an overall constant per linearly independent solution. A consistent choice for $f^{(l)}(p_{i}x)$ and $ g^{(l)}(p_{i}x)$ is 
\begin{equation}
f^{(l)}(p_{i}x) \equiv x j_{l}(p_{i}x), \  g^{(l)}(p_{i}x) \equiv i x h^{(1)}_{l}(p_{i}x),
\label{equ:choice}
\end{equation}
where $j_{l}(p_{i}x)$ is the spherical Bessel function and $h^{(l)}_{l}(p_{i}x)$ is the spherical Hankel function of the first kind. 

To keep track of the linearly independent solutions, we introduce a new subscript,
\begin{equation}
\chi_{in}(x) = \alpha_{i n}(x) f(p_{i} x) - \beta_{i n}(x) g(p_{i}x) 
\label{equ:subscript}
\end{equation} 
where we have dropped the angular momentum label $l$ for brevity, $n=1,2$ for the two independent solutions for a given $i=1,2$, which labels the particle state. Defining
\begin{equation}
\begin{aligned}
\boldsymbol{f}(x) \equiv \begin{bmatrix} f(ax) & 0 \\ 0 & f(cx) \end{bmatrix}, \ 
\boldsymbol{g}(x) \equiv \begin{bmatrix} g(ax) & 0 \\ 0 & g(cx) \end{bmatrix}, \ 
\boldsymbol{\alpha}(x) \equiv \begin{bmatrix} \alpha_{11}(x) &  \alpha_{12}(x) \\  \alpha_{21}(x) &  \alpha_{22}(x) \end{bmatrix}, \\
\boldsymbol{\beta}(x) \equiv \begin{bmatrix}  \beta_{11}(x) &  \beta_{12}(x) \\  \beta_{21}(x) &  \beta_{22}(x) \end{bmatrix}, \ 
\boldsymbol{\chi}(x) \equiv \begin{bmatrix} \chi_{11}(x) &  \chi_{12}(x) \\  \chi_{21}(x) &  \chi_{22}(x) \end{bmatrix} \ \ \ \ \ \ \ \ \ \ \ \ \ \ \ \ \ \
\end{aligned},
\label{equ:matrix_definition}
\end{equation}
we can rewrite equation (\ref{equ:guess}) in a compact form,
\begin{equation}
\boldsymbol{\chi}(x) = \boldsymbol{f}(x) \boldsymbol{\alpha}(x) - \boldsymbol{g}(x) \boldsymbol{\beta}(x).
\label{equ:matrix_equation}
\end{equation}
Further defining
\begin{equation}
\boldsymbol{\xi}(x) \equiv \boldsymbol{\chi}(x) \boldsymbol{\alpha^{-1}}(x), \ \boldsymbol{M}(x) \equiv  \boldsymbol{\beta}(x) \boldsymbol{\alpha^{-1}}(x),
\label{equ:matrix_parameters}
\end{equation}
we have
\begin{equation}
\boldsymbol{\xi}(x) = \boldsymbol{f}(x) - \boldsymbol{g}(x) \boldsymbol{M}(x).
\label{equ:matrix_guess}
\end{equation}

Taking the $x \rightarrow \infty$ limit of the choices for $f(p_{i}x)$ and $g(p_{i}x)$, we can see the virtue of the conventions and definitions employed so far,
\begin{equation} 
\lim\limits_{ x \rightarrow \infty} f(p_{i} x)= \frac{(-i)^{l+1}e^{ip_{i}x}+(i)^{l+1}e^{-ip_{i}x}}{2}, \;   \lim\limits_{ x \rightarrow \infty}g( p_{i} x) = (-i)^{l+2}e^{ip_{i}x}.
\label{equ:limiting_form}
\end{equation}
Inserting (\ref{equ:limiting_form}) into (\ref{equ:matrix_guess}) and comparing with equation (\ref{equ:spherical_waves}), one can find that the components of $\boldsymbol{M}$ are related to the scattering amplitudes as  
\begin{equation}
\frac{M_{11}(x \rightarrow \infty)}{a} = \alpha_{x} m_{x} \mathcal{F}_{x}, \ \ \ \frac{M_{21}(x \rightarrow \infty)}{a} = \alpha_{x} m_{x} \mathcal{F}_{y}
\label{equ:scattering_amp1}
\end{equation}
where the incoming scatterers are of type 1.
Similarly,
\begin{equation}
\frac{M_{12}(x \rightarrow \infty)}{c} = \alpha_{x} m_{x} \mathcal{F}_{x}, \ \ \ \frac{M_{22}(x \rightarrow \infty)}{c} = \alpha_{x} m_{x} \mathcal{F}_{y}
\label{equ:scattering_amp2}
\end{equation}
for incoming particles of type 2.
The $i^{th}$ column of $\boldsymbol{\xi}$ is interpreted as the scattered wave functions for the two particle states where the incoming states are of type $i$.
Next we make the definition,
\begin{equation}
U_{ij}(x) \equiv f(p_{i}x) g(p_{i} x) \delta_{ij} - g(p_{j}x) M_{ij}(x) g(p_{j}x).
\label{equ:u_definition}
\end{equation}
Using the formalism developed in this subsection, we can derive the following first-order different equation for $U_{ij}(x)$,
\begin{equation}
\frac{dU_{ij}(x)}{dx} = p_{i} \delta_{ij} + \left(p_{i} \frac{g'(p_{i}x)}{g(p_{i}x)}+p_{j} \frac{g'(p_{j}x)}{g(p_{j}x)}\right) U_{ij}(x) - U_{il}(x)\frac{\tilde{V}_{lm}(x)}{p_{l}}U_{mj}(x)
\label{equ:u_equation}
\end{equation}
where $\tilde{V}(x) \equiv \begin{bmatrix} 0 & -\frac{e^{-x / b}}{x} \\ -\frac{e^{-x / b}}{x} & 0 \end{bmatrix}$ and $g'(p_{i}x) \equiv {dg(p_{i}x)}/{d(p_{i}x)}$. As $x \rightarrow 0$, $\beta_{ij}(x) \rightarrow 0$ since the solution $\chi_{ij}(x)$ must be regular at the origin and we take $\alpha_{ij}(x) \rightarrow \delta_{ij}$.
Therefore, $M_{ij}(x \rightarrow 0) = 0$ and the initial condition for $U_{ij}(x)$ becomes,
\begin{equation}
U_{ij}(x \rightarrow 0) = f(p_{i} x) g(p_{i}x) \delta_{ij}. 
\label{equ:initial_condition}
\end{equation}
The advantage of this differential equation is that only logarithmic derivatives of the free solutions enter into the equation greatly increasing its numerical stability. We can now solve equation (\ref{equ:u_equation}) using the initial condition (\ref{equ:initial_condition}) for $U_{ij}(x)$. Once $U_{ij}(x)$ is known then the scattering amplitudes $\sim M_{ij}(x)$ can be obtained using the definition (\ref{equ:u_definition}). Finally, the scattering cross section can be calculated using equation (\ref{equ:diff_xsection}). It is useful to note that the transformation $\Delta m\rightarrow - \Delta m$ changes the incoming particles from one type to the other ($a \leftrightarrow c$). Therefore we only need to solve for $M_{11}(x)$ and $M_{21}(x)$, once $\Delta m$ is changed to $- \Delta m$, in order to obtain all scattering cross sections.
Also, there is an equation for $\alpha_{ij}(x)$ ($\beta_{ij}(x)$) which carries information needed to solve for the Sommerfeld enhancements but are not needed in calculating the self-scattering cross sections~\cite{Blennow:2016gde}.

\subsection{Formulae}

Here we develop formula for the total scattering cross section as well as the viscosity and transfer cross sections. 
Starting from equation (\ref{equ:diff_xsection}), we can write the expression for the total cross section as 
\begin{equation}
\sigma_{\rm tot} = \frac{p_{\rm{out}}}{p_{\rm{in}}} \sum_{l=0}^{\infty} \sum_{l'=0}^{\infty} (2l+1)(2l'+1)\mathcal{F}_{l} \mathcal{F}_{l'}^{*} \int d\Omega P_{l}(\cos\theta) P_{\it l'}^{*}(\cos\theta)).
\label{equ:xsection}
\end{equation}
Using the identity,
\begin{equation}
\int_{-1}^{1} dx P_{l}(x)  P_{l'}^{*}(x) = \frac{2\delta_{ll'}}{(2l+1)}
\label{equ:identity}
\end{equation}
the total cross section is given by 
\begin{equation}
\sigma_{\rm tot} = 4 \pi \frac{p_{\rm{out}}}{p_{\rm{in}}} \sum_{l=0}^{\infty} (2l+1) \left|\mathcal{F}_{l}\right|^{2}.
\label{equ:total_xsection}
\end{equation}
The transfer cross section is weighted such that forward scattering events (scattering angle $\theta \rightarrow 0$) do not contribute at all and backward scattering events ($\theta \rightarrow \pi$) give the largest contribution to the cross section,
\begin{equation}
\sigma_{T} \equiv \frac{p_{\rm{out}}}{p_{\rm{in}}} \int d\Omega \left| \sum_{l=0}^{\infty} (2l+1) P_{l}(\rm{cos}\theta) \mathcal{F}_{l} \right|^{2} (1-\rm{cos}\theta).
\label{equ:transfer_definition}
\end{equation} 
Using (\ref{equ:identity}) and the recursion relation  
\begin{equation}
(l+1)P_{(l+1)}(x) = (2l+1)xP_{l}(x) - l P_{(l-1)}(x)
\label{equ:recursion}
\end{equation}
we have
\begin{equation}
\int_{-1}^{1} dx P_{l}(x) P_{l'}^{*}(x) (1-x) = \frac{2}{(2l+1)} \left[\delta_{ll'} - \frac{(l+1)}{(2l+3)} \delta_{(l+1)l'} - \frac{l}{(2l-1)} \delta_{(l-1)l'}\right].
\label{equ:derived_identity_transfer}
\end{equation}
Identity (\ref{equ:derived_identity_transfer}) allows the transfer cross section to be written as,
\begin{equation}
\begin{aligned}
\sigma_{T} = 4 \pi \frac{p_{\rm{out}}}{p_{\rm{in}}} \sum_{l=0}^{\infty}  \left[ (2l+1) \left|\mathcal{F}_{l}\right|^{2} - (l+1) \mathcal{F}_{l} \mathcal{F}_{(l+1)}^{*} - l \mathcal{F}_{l} \mathcal{F}_{(l-1)}^{*}\right] \\
= 4 \pi \frac{p_{\rm{out}}}{p_{\rm{in}}} \sum_{l=0}^{\infty}  \left[ (2l+1) \left|\mathcal{F}_{l}\right|^{2} - 2 (l+1) Re( \mathcal{F}_{l} \mathcal{F}_{(l+1)}^{*} )\right].
\end{aligned}
\label{equ:long_transfer_xsection}
\end{equation}
We further write the scattering amplitude as a general complex number $\mathcal{F}_{l} \equiv \left| \mathcal{F}_{l} \right| e^{i \delta_{l}}$ and insert it into (\ref{equ:long_transfer_xsection}),
\begin{equation}
\sigma_{T} = 4 \pi \frac{p_{\rm{ou}t}}{p_{\rm{in}}} \sum_{l=0}^{\infty} (l+1) \left[ \left|\mathcal{F}_{(l+1)}\right|^{2} +  \left|\mathcal{F}_{l}\right|^{2} - 2 \left| \mathcal{F}_{(l+1)} \right| \left| \mathcal{F}_{l} \right| \rm{cos}(\delta_{(l+1)} - \delta_{l})\right].
\label{equ:transfer_xsection}
\end{equation}
Equation (\ref{equ:transfer_xsection}) has the benefit of being positive definite term-wise such that the sum is monotonically increasing. This property allows the sum to converge more quickly. The viscosity cross section is defined such that neither forward nor backward scattering contribute to the cross section,
\begin{equation}
\begin{aligned}
\sigma_{V} \equiv \frac{p_{\rm{out}}}{p_{\rm{in}}} \int d\Omega \left| \sum_{l=0}^{\infty} (2l+1) P_{l}(\rm{cos}\theta) \mathcal{F}_{l} \right|^{2} \rm{sin}^{2}\theta .
\label{equ:viscosity_definition}
\end{aligned}
\end{equation}
Following a similar procedure to the transfer cross section calculation and using identities (\ref{equ:identity}) and (\ref{equ:recursion}), we can derive the following identity
\begin{equation}
\begin{aligned}
 \int_{-1}^{1} dx P_{l}(x) P_{l'}^{*}(x) (1-x^{2}) =  \frac{2}{(2l+1)} \ \ \ \ \ \ \ \ \ \ \ \ \ \ \ \ \ \ \ \ \ \ \ \ \ \ \ \ \ \ \ \ \ \ \ \ \ \ \ \ \ \ \ \  \ \ \ \\ \left[ \left(1-\frac{(l+1)^{2}}{(2l+1)(2l+3)} - \frac{l^{2}}{(2l+1)(2l-1)}\right) \delta_{ll'} - \frac{(l+1)(l+2)}{(2l+3)(2l+5)} \delta_{(l+2)l'} - \frac{l(l-1)}{(2l-1)(2l-3)} \delta_{(l-2)l'} \right]. 
 \\ \\ 
\end{aligned}
\label{equ:derived_identity_viscosity}
\end{equation}
The viscosity cross section is then,
\begin{equation}
\begin{aligned}
\sigma_{V} = 4 \pi \frac{p_{\rm{out}}}{p_{\rm{in}}} \sum_{l=0}^{\infty} \left[ ((2l+1) - \frac{(l+1)^{2}}{(2l+3)} - \frac{l^{2}}{(2l-1)} ) \left| \mathcal{F}_{l} \right|^{2} - \frac{(l+1)(l+2)}{(2l+3)} \mathcal{F}_{l} \mathcal{F}_{(l+2)}^{*} - \frac{l(l-1)}{(2l-1)} \mathcal{F}_{l} \mathcal{F}_{(l-2)}^{*} \right] \\
= 4 \pi \frac{p_{\rm{out}}}{p_{\rm{in}}} \sum_{l=0}^{\infty} \left[ \frac{2(2l+1)(l^{2}+l-1)}{(2l+3)(2l-1)} \left|\mathcal{F}_{l} \right|^{2} - \frac{2(l+2)(l+1)}{(2l+3)} Re(\mathcal{F}_{l} \mathcal{F}_{(l+2)}^{*}) \right]. \ \ \ \ \ \ \ \ \ \ \ \ \ \ \ \ \ \ \ \ \ \ \ \ \ 
\end{aligned}
\label{equ:long_viscosity_xsection}
\end{equation}
Rewriting $\mathcal{F}_{l}$ in polar form gives the final result,
\begin{equation}
\sigma_{V} = 4 \pi \frac{p_{\rm{out}}}{p_{\rm{in}}} \sum_{l=0}^{\infty} \frac{(l+1)(l+2)}{(2l+3)} \left[\left| \mathcal{F}_{(l+2)} \right|^{2}+\left| \mathcal{F}_{l} \right|^{2} - 2 \left| \mathcal{F}_{(l+2)} \right| \left| \mathcal{F}_{l}\right| \rm{cos}(\delta_{(l+2)} - \delta_{l})\right]
\label{equ:viscosity_xsection}
\end{equation}
which is again term-wise positive definite. 

\acknowledgments
We thank Stefan Clementz for many valuable conversations which were important to the development of the numerical method in this work, Yiming Zhong for providing the data of the core-collapse constraints, and Yue Zhang for useful discussions. This work was supported by the U. S. Department of Energy under Grant No. de-sc0008541 (HBY) and UCR Academic Senate Regents Faculty Development Award (HBY).

\bibliography{Bibliography}

\end{document}